\documentclass[a4paper,showpacs,aps,prd,nofootinbib]{revtex4}
\usepackage[latin9]{inputenc}
\usepackage{hyperref}
\hypersetup{
  colorlinks=true,        
  linkcolor=blue,         
  citecolor=cyan,         
}
\usepackage{breakurl}
\usepackage{graphicx}
\usepackage{epsf}
\usepackage{epsfig}
\usepackage{amssymb,amsmath}
\usepackage[usenames]{color}
\usepackage{amssymb}
\usepackage{times}

\newcommand{\ie}{i.e.,~}



\begin{document}

\title{1+1 dimensional relativistic magnetohydrodynamics with longitudinal acceleration}

\author{Duan She$^{~1,2}$}
\author{Ze Fang Jiang$^{~1,2,3}$}
\author{Defu Hou$^{~1,2}$}
\author{C.~B.~Yang$^{~1,2}$}

\affiliation{$^1$ Key Laboratory of Quark and Lepton Physics, Ministry of Education, Wuhan, 430079, China}
\affiliation{$^2$ Institute of Particle Physics, Central China Normal University, Wuhan 430079, China}
\affiliation{$^3$ Department of Physics and Electronic-information Engineering, Hubei Engineering University,~Xiaogan 432000,~China}

\begin{abstract}
Non-central heavy-ion collisions generate the strongest magnetic field of the order of $10^{18}-10^{19}$ Gauss due to the electric current produced by the positively charged spectators that travel at nearly the speed of light. Such transient electromagnetic fields may induce  various novel effects in the hydrodynamic description of the quark gluon plasma for non-central heavy-ion collisions. We investigate  the longitudinal acceleration effects on the 1+1 dimensional relativistic magnetohydrodynamics with transverse magnetic fields. We analyze  the proper time evolution of the system energy density. We find  that the  longitudinal acceleration parameter $\lambda^*$, magnetic field decay parameter $a$, equation of state $\kappa$, and initial magnetization $\sigma_0$ have nontrivial effects on the evolutions of the system energy density and  temperature.
\end{abstract}


\maketitle

\section{Introduction}
The heavy ion collisions have the advantage of being able to study the hot medium in detail under controlled environments.
It has recently been reported that the strongest magnetic fields of the order of $10^{18}-10^{19}$ Gauss are produced in non-central heavy-ion collisions by the electric current from the positively charged spectators that travel at nearly the speed of light ~\cite{Bzdak2012,Deng2012,Kharzeev2008}.
It is expected that such a huge magnetic field may have important consequences on the dynamics of the quark-gluon matter produced in heavy-ion collisions ~\cite{Gyulassy2005}.
In particular, it has been proposed that the interplay of quantum anomalies with ultra-intense magnetic field results in several special transport phenomena
that are closely related to chiral anomaly and thus are called anomalous transports ~\cite{Xu-Guang Huang2016,Kharzeev2016}.
The external magnetic fields may induce charge separation in a chirality-imbalanced medium, namely the chiral magnetic effect (CME) ~\cite{Kharzeev2008,Fukushima2008}
which has been observed at RHIC and the LHC and the measurements signals indeed consistent with the predictions of the CME
~\cite{STAR2010,STAR2009,ALICE2016}.
Along with CME, the chiral separation effect (CSE) ~\cite{Metlitski2005,Son2004} represents the generation of the axial current along the external magnetic field in the presence of finite vector charge density. The duality between CME and CSE leads to the interesting collective effect,
called ``chiral magnetic wave"(CMW) ~\cite{Kharzeev2011}, which induces a quadrupole deformation of the electric charge distribution
that might be responsible for breaking the degeneracy between the elliptic flows of $\pi^{\pm}$ ~\cite{Burnier2011}.

The relativistic hydrodynamic models have so far nicely described the thermodynamic evolution of the produced matter and the
experimentally measured flow harmonics in heavy-ion collisions ~\cite{Romatschke2007,Luzum2008,Songprl2008,Songprc2008,Schenke2012,
Roy2012,Niemi2012}. The influence of strong magnetic fields on the hot and dense nuclear matter have been intensively investigated.
In principle, such studies can be accomplished by solving the relativistic magnetohydrodynamics (MHD) equations that takes into account
the dynamical coupling of the magnetic field to the fluid.
Although the magnetic field generated in heavy-ion collisions rapidly decays in the vacuum,
the hot medium created in the heavy-ion collision as a conducting plasma might substantially
delay the decay of the magnetic field through the generation of an induction current due to
Lenz's law ~\cite{Gursoy2014,Zakharov2014,Tuchin2013}.

In Refs.~\cite{Roy2015}, one-dimensional magnetic fluid had been investigated by using the longitudinally boost-invariant Bjorken flow~\cite{Bjorken1983} with a transverse and time-dependent homogeneous magnetic field.
In ideal MHD limits, with the infinite electrical conductivity and neglecting other dissipative effects such as viscosity and thermal conduction, it is extraordinary that the evolution of energy density is the same as the case without magnetic fields due to ``frozen-flux theorem".
Later, a nonzero magnetization effect is introduced to the Bjorken flow in MHD~\cite{Shi2016}.

QGP expanding in the beam direction can be described by considering a boost invariant Bjorken flow, that is known to be a good approximation at mid-rapidity. It leads to a flat rapidity distribution of final particle,
which is inconsistent with observations at RHIC, except for a narrow region around mid-rapidity ~\cite{Bjorken1983}.
It has been pointed out that in realistic situations the energy density at mid-rapidity decreases faster than in the Bjorken flow.
Although the Bjorken-estimation for the initial energy density is widely used, the longitudinal expansion dynamics of hydrodynamics seems ~\cite{Csorgo:2006ax,Csand:2016arx,Jiang2017,Jiang2018} to be able to offer a more realistic estimation for the initial energy density estimation and the final state description. Acceleration effects are important in the estimation of the initial energy density even at mid-rapidity, if the expanding system is finite: even the most central fluid element exert a force on the volume elements closer to the surface, and this work decreases the internal energy of cells even at mid-rapidity.

This paper is organized as follows. In Sec.\ref{section2}, the ideal-MHD framework with acceleration effects is presented.
In Sec.\ref{section3}, we present the evolution of the energy density in ideal transverse MHD with longitudinal expansion dynamics.
We consider the decay of the energy density within an external homogeneous magnetic field which decays with a power-law in proper time.
In subsection.~\ref{CNC approximation}, an exact analytic solution under the CNC approximation is obtained.
In subsection.~\ref{numerical solution}, we show the results obtained from numerical method for a realistic equation of state (EoS).
Finally, we discuss and conclude in the last section.
Throughout this work, $u^{\mu}=\gamma\left(1,\overrightarrow{\boldsymbol{v}}\right)$
is the four-velocity field that satisfies $u^{\mu}u_{\mu}=1$ and
the spatial projection operator $\Delta^{\mu\nu}=g^{\mu\nu}-u^{\mu}u^{\nu}$ is defined with
the Minkowski metric $g^{\mu\nu}={\rm diag}\left(1,-1,-1,-1\right)$.
It is note-worthy that the orthogonality relation $\Delta^{\mu\nu}u_{\nu}=0$ is satisfied.
We adopt the standard convention for the summation over repeated indices.

\section{ideal MHD with acceleration}
\label{section2}

Relativistic magnetohydrodynamics (RMHD for short) concerns the mutual interaction of fluid flow and magnetic fields.
The fluids in question must be electrically conducting and non-magnetic.
The RMHD evolution equations describe the dynamics of the overall system based on local conservation of this fluid current
(associated to the net-baryon current or to any other conserved charge)
and the total (matter and fields) energy-momentum as well as on the additional assumption of local thermal equilibrium.

Consider an non-viscous fluid coupling with  a magnetic field.
We assume that the medium is perfectly conducting and the electric field in the co-moving frame vanishes
to avoid the onset of huge currents in the plasma.
The total energy-momentum tensor of ideal fluid is given by ~\cite{Huang2010,Giacomazzo2006,Giacomazzo2007}
\begin{eqnarray}
T^{\mu\nu}=(e+p+B^{2})u^{\mu}u^{\nu}-\left(p+\frac{B^{2}}{2}\right)g^{\mu\nu}-B^{\mu}B^{\nu},
\label{1}
\end{eqnarray}
where
\begin{eqnarray}
B^{2}=B^{\mu}B_{\mu},\quad B^{\mu}=\frac{1}{2}\epsilon^{\mu\nu\alpha\beta}u_{\nu}F_{\alpha\beta},
\label{2}
\end{eqnarray}
here $e,~p$ and $F_{\alpha\beta}$ are the fluid energy density, pressure and the Faraday tensor.
Here, $\epsilon^{\mu\nu\alpha\beta}$ is the completely antisymmetric four tensor with $\epsilon^{0123}=-\epsilon_{0123}=1$.
The magnetic field four-vector $B^{\mu}$ is a space-like vector with modulus $B^{\mu}B_{\mu}=-B^{2}$ and is orthogonal to $u^{\mu}$, \ie $B^{\mu}u_{\mu}=0$, where $B=|\vec{\boldsymbol{B}}|$ and $\vec{\boldsymbol{B}}$
is the magnetic field three-vector in the frame moving with four-velocity $u^{\mu}$.

In the present paper we consider the special case of a fluid flow with the external magnetic field
$\vec{\boldsymbol{B}}$ directed along the transverse plane.
This setup is consistent with the scenario in non-central heavy ion collision at top RHIC energy~\cite{Deng2012}.
The system of ideal RMHD equations can be closed by choosing the rather general EoS
\begin{equation}
p={c}_{s}^{2}\,e=\frac{1}{\kappa}\,e\,,
\label{3}
\end{equation}
where $c_s$ stands for the local speed of sound which is assumed to be a constant.
In a fully realistic solution, we should use results form the lattice QCD, with the speed of sound being a function of temperature~\cite{Gupta2004,Qin2015}.
However, in current work we approximate $c_s(T)$ as a temperature independent constant $c_s$.
We postpone the analysis of the case of $c_s(T)$ for a later, more detailed investigation.

We decompose the covariant derivative as
\begin{eqnarray}
\partial_{\mu}=u_{\mu}D+\nabla_{\mu},
\label{4}
\end{eqnarray}
where $D=u^{\mu}\partial_{\mu}$ indicates the time derivative in the local rest frame,
and $\nabla^{\mu}=\Delta^{\mu\nu}\partial_{\nu}$ is the spatial gradient in the local rest frame.
The energy conservation equation is derived by projecting the conservation law $\partial_{\mu}T^{\mu\nu}=0$ along the fluid four-velocity $u^{\mu}$,
\begin{eqnarray}
u_{\mu}\partial_{\nu}T^{\mu\nu}	&=&	u_{\mu}u^{\mu}u^{\nu}\partial_{\nu}(e+p+B^{2})+(e+p+B^{2})u_{\mu}\partial_{\nu}(u^{\mu}u^{\nu})-u_{\mu}\partial_{\nu}\left[\left(p+\frac{B^{2}}{2}\right)g^{\mu\nu}\right]-u_{\mu}\partial_{\nu}(B^{\mu}B^{\nu})\nonumber\\
	&=&	u^{\nu}\partial_{\nu}(e+p+B^{2})+(e+p+B^{2})\partial_{\nu}u^{\nu}+0-u^{\nu}\partial_{\nu}\left(p+\frac{B^{2}}{2}\right)-\partial_{\nu}(u_{\mu}B^{\mu}B^{\nu})+B^{\mu}B^{\nu}\partial_{\nu}u_{\mu}
	\nonumber\\
	&=&		D(e+p+B^{2})+(e+p+B^{2})\theta-D\left(p+\frac{B^{2}}{2}\right)
	\nonumber\\
	&=&		D\left(e+\frac{B^{2}}{2}\right)+(e+p+B^{2})\theta
	\nonumber\\
	&=&		0,
\label{5}
\end{eqnarray}
where $\theta\equiv\partial_{\mu}u^{\mu}=\nabla_{\mu}u^{\mu}$ is the expansion factor
and we have used relation $u^{\mu}B_{\mu}=0$ and $B^{\nu}\partial_{\nu}u_{\mu}=0$,
since $u_{\mu}=\left(u_{0},0,0,u_{z}\right)$ and $B_{\mu}=\left(0,B_{x},B_{y},0\right)$ in our setup.
Thus one obtains the energy-conservation equation as follows,
\begin{eqnarray}
D\left(e+\frac{B^{2}}{2}\right)+(e+p+B^{2})\theta=0 \,.
\label{6}
\end{eqnarray}

The relativistic version of the MHD Euler equation is retrieved by projecting the energy-momentum conservation equation onto the direction orthogonal to $u^{\mu}$,
\begin{eqnarray}
\triangle_{\mu\nu}\partial_{\alpha}T^{\alpha\nu}	&=&		(e+p+B^{2})\triangle_{\mu\nu}\partial_{\alpha}(u^{\alpha}u^{\nu})+0-\triangle_{\mu\nu}\partial^{\nu}\left(p+\frac{B^{2}}{2}\right)-\triangle_{\mu\nu}\partial_{\alpha}(B^{\alpha}B^{\nu})\nonumber\\
	&=&		(e+p+B^{2})(g_{\mu\nu}-u_{\mu}u_{\nu})\partial_{\alpha}(u^{\alpha}u^{\nu})-\triangle_{\mu\nu}\partial^{\nu}\left(p+\frac{B^{2}}{2}\right)-(g_{\mu\nu}-u_{\mu}u_{\nu})\partial_{\alpha}(B^{\alpha}B^{\nu})\nonumber\\
	&=&		(e+p+B^{2})\left[u^{\alpha}\partial_{\alpha}u_{\mu}+u_{\mu}\partial_{\alpha}u^{\alpha}-u_{\mu}u_{\nu}u^{\alpha}\partial^{\alpha}u^{\nu}-u_{\mu}u_{\nu}u^{\nu}\partial_{\alpha}u^{\alpha}\right]-\triangle_{\mu\nu}\partial^{\nu}\left(p+\frac{B^{2}}{2}\right)\nonumber\\
&&-B^{\alpha}\partial_{\alpha}B_{\mu}-B_{\mu}\partial_{\alpha}B^{\alpha}+u_{\mu}u_{\nu}B^{\alpha}\partial_{\alpha}B^{\nu}+u_{\mu}u_{\nu}B^{\nu}\partial_{\alpha}B^{\alpha}
\nonumber\\
	&=&		(e+p+B^{2})Du_{\mu}-\triangle_{\mu\nu}\partial^{\nu}\left(p+\frac{B^{2}}{2}\right)-B^{\alpha}\partial_{\alpha}B_{\mu}-B_{\mu}\partial_{\alpha}B^{\alpha}-u_{\mu}B^{\alpha}B^{\nu}\partial_{\alpha}u_{\nu}\nonumber\\
&=&		0.
\label{7}
\end{eqnarray}

The last three terms vanish and leads to the Euler equation as follow,
\begin{equation}
(e+p+B^{2})Du_{\mu}-\nabla_{\mu}\left(p+\frac{B^{2}}{2}\right)=0\,.
\label{8}
\end{equation}

We use the well-known Rindler coordinates $\tau=\sqrt{t^{2}-r^{2}}$ and $\eta_{s}=\frac{1}{2}{\rm log}\left(\left(t+r\right)/\left(t-r\right)\right)$ as independent variables inside the forward lightcone and parametrize the fluid velocity as $v={\rm tanh}\Omega$, and the fluid rapidity $\Omega$ depends only on $\eta_s$ here. (For simplicity, we will use $\Omega$ to denotes $\Omega(\eta_s)$, and $\Omega^{\prime}$ denotes $d\Omega/d\eta_{s}$). One obtains
\begin{eqnarray}
D&=&u^{\mu}\partial_{\mu}=u^{0}\partial_{t}+u^{z}\partial_{z}\nonumber\\
&=&{\rm cosh}\Omega\left({\rm cosh}\eta_{s}\frac{\partial}{\partial\tau}-\frac{{\rm sinh}\eta_{s}}{\tau}\frac{\partial}{\partial\eta_{s}}\right)+{\rm sinh}\Omega\left(-{\rm sinh}\eta_{s}\frac{\partial}{\partial\tau}+\frac{{\rm cosh}\eta_{s}}{\tau}\frac{\partial}{\partial\eta_{s}}\right)\nonumber\\
&=&{\rm cosh}(\Omega-\eta_{s})\frac{\partial}{\partial\tau}+\frac{1}{\tau}{\rm sinh}(\Omega-\eta_{s})\frac{\partial}{\partial\eta_{s}}
\label{9}
 \end{eqnarray}
and
\begin{eqnarray}
\theta&=&\partial_{\mu}u^{\mu}=\partial_{t}u^{0}+\partial_{z}u^{z}\nonumber\\
&=&\left({\rm cosh}\eta_{s}\frac{\partial}{\partial\tau}-\frac{{\rm sinh}\eta_{s}}{\tau}\frac{\partial}{\partial\eta_{s}}\right){\rm cosh}\Omega+\left(-{\rm sinh}\eta_{s}\frac{\partial}{\partial\tau}+\frac{{\rm cosh}\eta_{s}}{\tau}\frac{\partial}{\partial\eta_{s}}\right){\rm sinh}\Omega\nonumber\\
&=&{\rm cosh}\eta_{s}{\rm sinh}\Omega\frac{\partial\Omega}{\partial\tau}-\frac{{\rm sinh}\eta_{s}}{\tau}{\rm sinh}\Omega\frac{\partial\Omega}{\partial\eta_{s}}-{\rm sinh}\eta_{s}{\rm cosh}\Omega\frac{\partial\Omega}{\partial\tau}+{\rm \frac{{\rm cosh}\eta_{s}}{\tau}{\rm cosh}\Omega}\frac{\partial\Omega}{\partial\eta_{s}}\nonumber\\
&=&{\rm sinh}(\Omega-\eta_{s})\frac{\partial\Omega}{\partial\tau}+\frac{1}{\tau}{\rm cosh}(\Omega-\eta_{s})\frac{\partial\Omega}{\partial\eta_{s}}.
\label{10}
 \end{eqnarray}

The peak value of the magnetic field $\vec{\boldsymbol{B}}$ is well determined by using event-by-event simulations with in the Monte-Carlo Glauber model~\cite{Bloczynski2013}.
Nevertheless, the lifetime of magnetic field is still an open question so far. We assume in this paper the homogeneous magnetic field obeys a power-law decay in proper time~\cite{Roy2015},
\begin{equation}
\overrightarrow{B}(\tau)=\overrightarrow{B}_{0}\left(\frac{\tau_{0}}{\tau}\right)^{a}.
\label{11}
\end{equation}
Here $a=1$ correspond to the ideal-MHD case~\cite{Davidson2017}, where $a>1$ corresponds to the case with the magnetic field decaying steeper than the ideal-MHD case,
and $a<1$ corresponds to a decay slower than in the ideal-MHD limit.
$\tau_0$ is the initial proper time of the fluid expansion and $B_{0}\equiv B(\tau_{0})$ is the initial magnetic field strength.

Above assumptions allow one to rewrite the conservation equations in Rindler coordinate as follows
\begin{eqnarray}
&&\tau\frac{\partial\widetilde{e}}{\partial\tau}+{\rm tanh}(\Omega-\eta_{s})\frac{\partial\widetilde{e}}{\partial\eta_{s}}+
\left(\widetilde{e}(1+c_{s}^{2})+\sigma_{0}\left(\frac{\tau_{0}}{\tau}\right)^{2a}\right)\Omega'=\sigma_{0}a\left(\frac{\tau_{0}}{\tau}\right)^{2a},~
\label{12}\\
&&\frac{\partial\widetilde{e}}{\partial\eta_{s}}={\rm tanh}(\Omega-\eta_{s})\left[\frac{\sigma_{0}}{c_{s}^{2}}
\left(\frac{\tau_{0}}{\tau}\right)^{2a}(a-\Omega')-\frac{1+c_{s}^{2}}{c_{s}^{2}}\widetilde{e}\Omega'-\tau\frac{\partial\widetilde{e}}{\partial\tau}\right],
\label{13}
\end{eqnarray}
with the dimensionless quantities $\widetilde{e}\equiv e/e_{0},~\sigma_{0}\equiv B_{0}^{2}/e_{0}$.

The combination of energy conservation equation Eq.~(\ref{12}) and the Euler equation Eq.~(\ref{13}) generates a partial differential equation,

\begin{eqnarray}
\tau\frac{\partial\widetilde{e}}{\partial\tau}=\left(\frac{{\rm sinh}^{2}(\Omega-\eta_{s})}{c_{s}^{2}}-{\rm cosh}^{2}(\Omega-\eta_{s})\right)\left[\left(\sigma_{0}\left(\frac{\tau_{0}}{\tau}\right)^{2a}+\widetilde{e}(1+c_{s}^{2})\right)\Omega'-\sigma_{0}a\left(\frac{\tau_{0}}{\tau}\right)^{2a}\right].
\label{14}
\end{eqnarray}
\section{energy-density evolution}
\label{section3}
The exact solution with CNC approximation (Sec.\ref{CNC approximation}) and numerical solution (Sec.\ref{numerical solution}) of energy density evolution in MHD are presented in this section step by step.

\subsection{Exact solution of MHD with CNC approximation}
\label{CNC approximation}
For a perfect fluid with longitudinal accelerating expansion, one finds $\Omega\neq\eta_{s}$.
The exact solution for such longitudinal accelerating hydrodynamics is the well-known CNC solution with $\Omega=\lambda\eta_{s}$ and $\kappa=1$, $c_{s}^{2}=\frac{1}{\kappa}=1,~\Omega'=\lambda,~\Omega''=0$. From Eq.(\ref{14}), one gets

\begin{eqnarray}
\tau\frac{\partial\widetilde{e}}{\partial\tau}=
\sigma_{0}a\left(\frac{\tau_{0}}{\tau}\right)^{2a}-\left(\sigma_{0}\left(\frac{\tau_{0}}{\tau}\right)^{2a}+2\widetilde{e}\right)\lambda.
\label{15}
\end{eqnarray}
The solution $\widetilde{e}(\tau,\eta_{s})$ is
\begin{eqnarray}
\widetilde{e}(\tau,\eta_{s})=-\frac{1}{2}\sigma_{0}\left(\frac{\tau_{0}}{\tau}\right)^{2a}+\tau^{-2\lambda}C(\eta_{s}),
\label{16}
\end{eqnarray}
where $C(\eta_{s})$ is an undetermined function related to the $\eta_s$ part of the energy density $\widetilde{e}(\tau,\eta_{s})$.

Putting Eq.(\ref{16}) into the Euler equation Eq.(\ref{13}), one gets
\begin{eqnarray}
C(\eta_{s})=C.
\label{17}
\end{eqnarray}
Then substituteing Eq.(\ref{17}) to Eq.(\ref{16}) and using the initial condition $\widetilde{e}_{0}(\tau_0,0)=1$, one obtains
\begin{eqnarray}
C=\frac{1}{2}(2+\sigma_{0})\tau_{0}^{2\lambda}.
\label{18}
\end{eqnarray}
Finally, inputting Eq.(\ref{18}) and Eq.(\ref{17}) into Eq.(\ref{16}), an analytical solution of the fluid energy density with CNC approximation can be written as follow
\begin{eqnarray}
\widetilde{e}(\tau,\eta_{s})=\left(\frac{\tau_{0}}{\tau}\right)^{2\lambda}+\frac{\sigma_{0}}{2}\left[\left(\frac{\tau_{0}}{\tau}\right)^{2\lambda}-\left(\frac{\tau_{0}}{\tau}\right)^{2a}\right].
\label{19}
\end{eqnarray}

\begin{table}[htb]
\begin{tabular}{|c|c|c|c|}
\hline
$$ & $a\to\lambda$ & $a\gg\lambda$ & $a\ll\lambda$
\\
\hline
$\widetilde{e}(\tau,\eta_{s})$ & $\left(\frac{\tau_{0}}{\tau}\right)^{2\lambda}$ & $\frac{2+\sigma_{0}}{2}\left(\frac{\tau_{0}}{\tau}\right)^{2\lambda}$ & $-\frac{\sigma_{0}}{2}\left(\frac{\tau_{0}}{\tau}\right)^{2a}$
\\
\hline
\end{tabular}
\caption{The fluid energy density in the three kinds of limit conditions}
\end{table}
Once again, it is possible to see that in the limit of vanishing magnetization $\sigma_{0}=0$ and $\lambda=1$, Eq.(\ref{19}) coincides with the solution for Bjorken flow.
If $\sigma_{0}=0$ and $\lambda \neq 1$, the solution coincides with CNC solution.
Furthermore, for $\sigma_{0}\neq 0$ and $\lambda=1$, one obtains the same solution as the Bjorken-Victor type flow~\cite{Roy2015}.

One can obtain the extreme value of the energy density from the following steps
\begin{eqnarray}
\frac{\partial\widetilde{e}(\tau,\eta_{s})}{\partial\tau}&&=\frac{a\sigma_{0}\left(\frac{\tau_{0}}{\tau}\right)^{2a}-\lambda(2+\sigma_{0})\left(\frac{\tau_{0}}{\tau}\right)^{2\lambda}}{\tau}=0
\label{20}\\
&&\Rightarrow\tau=\tau_{0}\left(\frac{a\sigma_{0}}{\lambda\left(2+\sigma_{0}\right)}\right)^{\frac{1}{2(a-\lambda)}}.
\label{21}
\end{eqnarray}
Unfortunately, these CNC solutions have a shortcoming, namely the acceleration parameter $\lambda$ becomes
a free fit parameter only for the superhard EoS of $\kappa=1,~e=p$.
In this case, the speed of sound is equal to the speed of light $c$, so the investigation was thought to be rather academic.

\subsection{Numerical solution for MHD}
\label{numerical solution}
To get a realistic solution of the energy density, we consider the case in which $\Omega\equiv\lambda\eta_{s}\equiv(1+\lambda^{*})\eta_{s}$ with $\lambda^{*}$ being a very small constant acceleration parameter ($0<\lambda^{*}\ll1$) and $\Omega'=1+\lambda^{*},~\Omega''=0$.

Thus, the energy equation and Euler equation can be expressed as
\begin{eqnarray}
&&\tau\frac{\partial\widetilde{e}}{\partial\tau}+{\rm tanh}(\lambda^{*}\eta_{s})\frac{\partial\widetilde{e}}{\partial\eta_{s}}+\left(\widetilde{e}(1+c_{s}^{2})+\sigma_{0}\left(\frac{\tau_{0}}{\tau}\right)^{2a}\right)(1+\lambda^{*})=\sigma_{0}a\left(\frac{\tau_{0}}{\tau}\right)^{2a},\label{22}\\
&&\frac{\partial\widetilde{e}}{\partial\eta_{s}}={\rm tanh}(\lambda^{*}\eta_{s})\left[\frac{\sigma_{0}}{c_{s}^{2}}\left(\frac{\tau_{0}}{\tau}\right)^{2a}(a-1-\lambda^{*})-\frac{1+c_{s}^{2}}{c_{s}^{2}}\widetilde{e}(1+\lambda^{*})-\tau\frac{\partial\widetilde{e}}{\partial\tau}\right].
\label{23}
\end{eqnarray}

The combination of energy equation Eq.(\ref{22}) and Euler equation Eq. (\ref{23}) can be rewritten as follows
\begin{eqnarray}
\tau\frac{\partial\widetilde{e}}{\partial\tau}&=&\left(\kappa{\rm sinh}^{2}(\lambda^{*}\eta_{s})-{\rm cosh}^{2}(\lambda^{*}\eta_{s})\right)\left[\left(\sigma_{0}\left(\frac{\tau_{0}}{\tau}\right)^{2a}+\widetilde{e}(1+\frac{1}{\kappa})\right)(1+\lambda^{*})-\sigma_{0}a\left(\frac{\tau_{0}}{\tau}\right)^{2a}\right],
\label{24}\\
\frac{\partial\widetilde{e}}{\partial\eta_{s}}&=&\frac{1}{2}{\rm sinh}(2\lambda^{*}\eta_{s})\left[\kappa\sigma_{0}\left(\frac{\tau_{0}}{\tau}\right)^{2a}(a-1-\lambda^{*})-(1+\kappa)\widetilde{e}(1+\lambda^{*})-\sigma_{0}a\left(\frac{\tau_{0}}{\tau}\right)^{2a}\right.\nonumber\\
&&\left.+\left(\widetilde{e}(1+\frac{1}{\kappa})+\sigma_{0}\left(\frac{\tau_{0}}{\tau}\right)^{2a}\right)(1+\lambda^{*})\right].
\label{25}
\end{eqnarray}
The main idea of solving the above partial differential equations is to treat this two PDEs as two ordinary differential equations with a given initial condition $\widetilde{e}(\tau_{0},0)=1$. Then one can get the relation between $\widetilde{e}$ and $\tau$ from Eq. (\ref{24}). The sets of data obtained above can be taken as the initial conditions for solving Eq.(\ref{25}). For this purpose, one obtains the full profile of energy density right away and obtains the evolution of temperature by using relation $e\propto T^{\kappa+1}$.

\begin{figure}[htb]
\begin{minipage}[htb]{8.5cm}
\centerline{\epsfig{figure=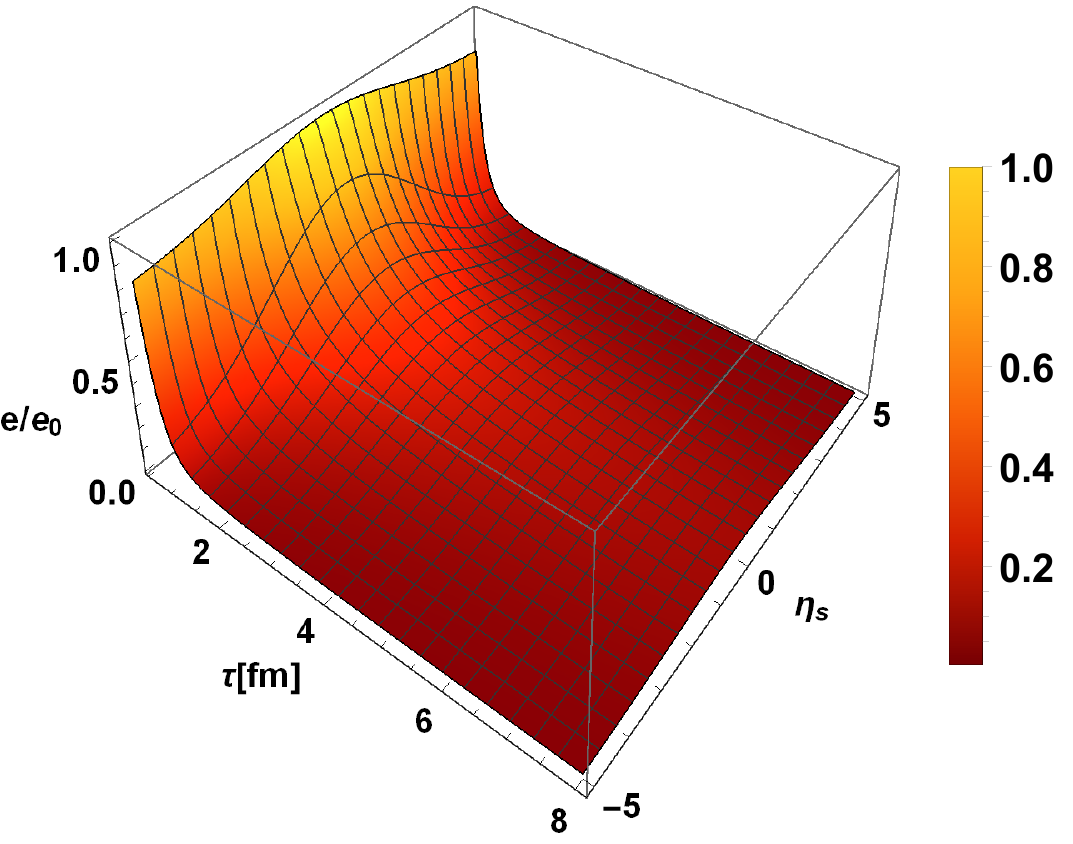,width=8cm}}
\end{minipage}
\hfill
\begin{minipage}[htb]{8.5cm}
\centerline{\epsfig{figure=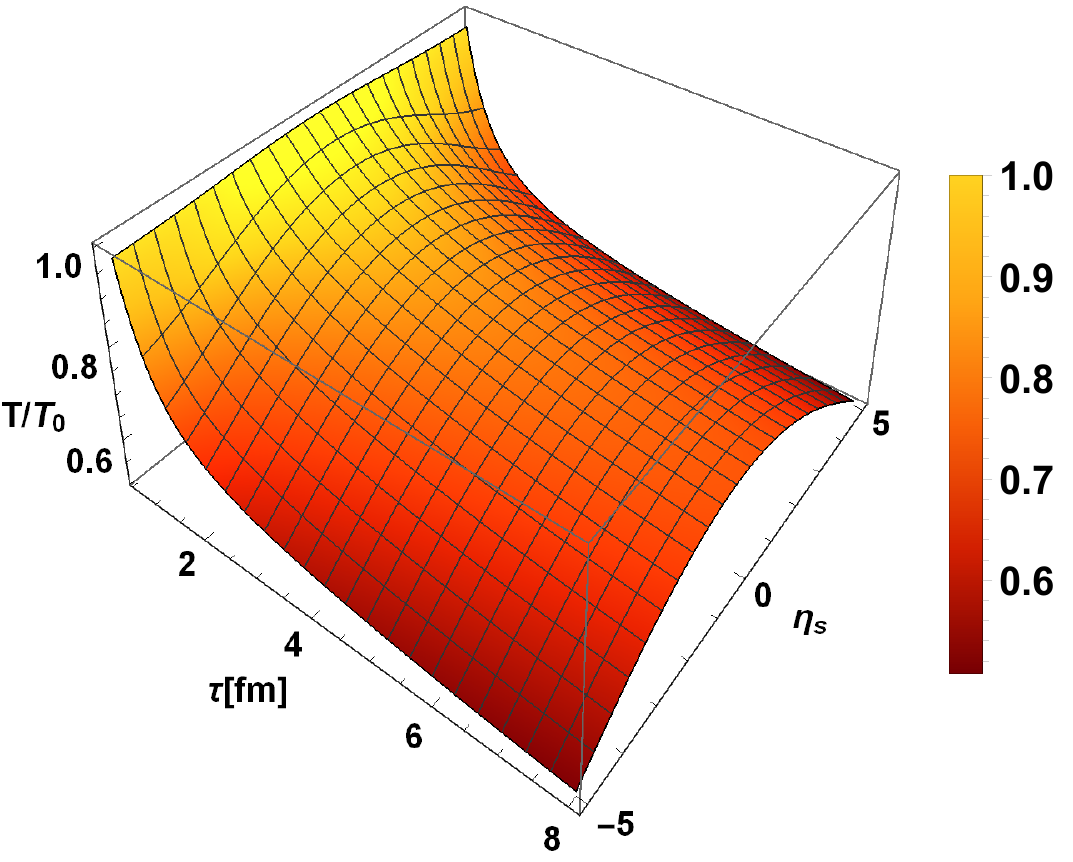,width=8cm}}
\end{minipage}
\caption{Left panel indicates the fluid energy density $e/e_{0}$, while the right panel shows the temperature $T/T_{0} $(right panel) profile, with parameters $a=2,\sigma_{0}=1.0,\kappa=7,\lambda^{*}=0.03$.}
\label{fig1}
\end{figure}

Fig. \ref{fig1} reports numerical solution of the fluid energy density and the temperature of accelerating fluid in one-dimensional relativistic magnetohydrodynamics with parameters $a=2,~\sigma_{0}=1.0,~\kappa=7,~\lambda^{*}=0.03$. The profile of $\widetilde{e}(\tau,~\eta_{s})$ is a (1+1) dimensional scaling solution, and it contains not only acceleration but also the magnetic field dependent terms now with the $\eta_s$ dependence of the Gaussian form.
Note that (1) if $\lambda^{*}=0$ and $\sigma_{0}=0$ one obtains the Bjorken solutions, (2) if $\lambda^{*}=0$ and $\sigma_0\neq0$ one obtains the same solution as the Bjorken-Victor type flow, (3) if $\lambda^{*}\neq 0$ and $\sigma_{0}=0$ one obtains the case of the well-known CNC solution.

\begin{figure}[htb]
\begin{minipage}[htb]{5cm}
\centerline{\epsfig{figure=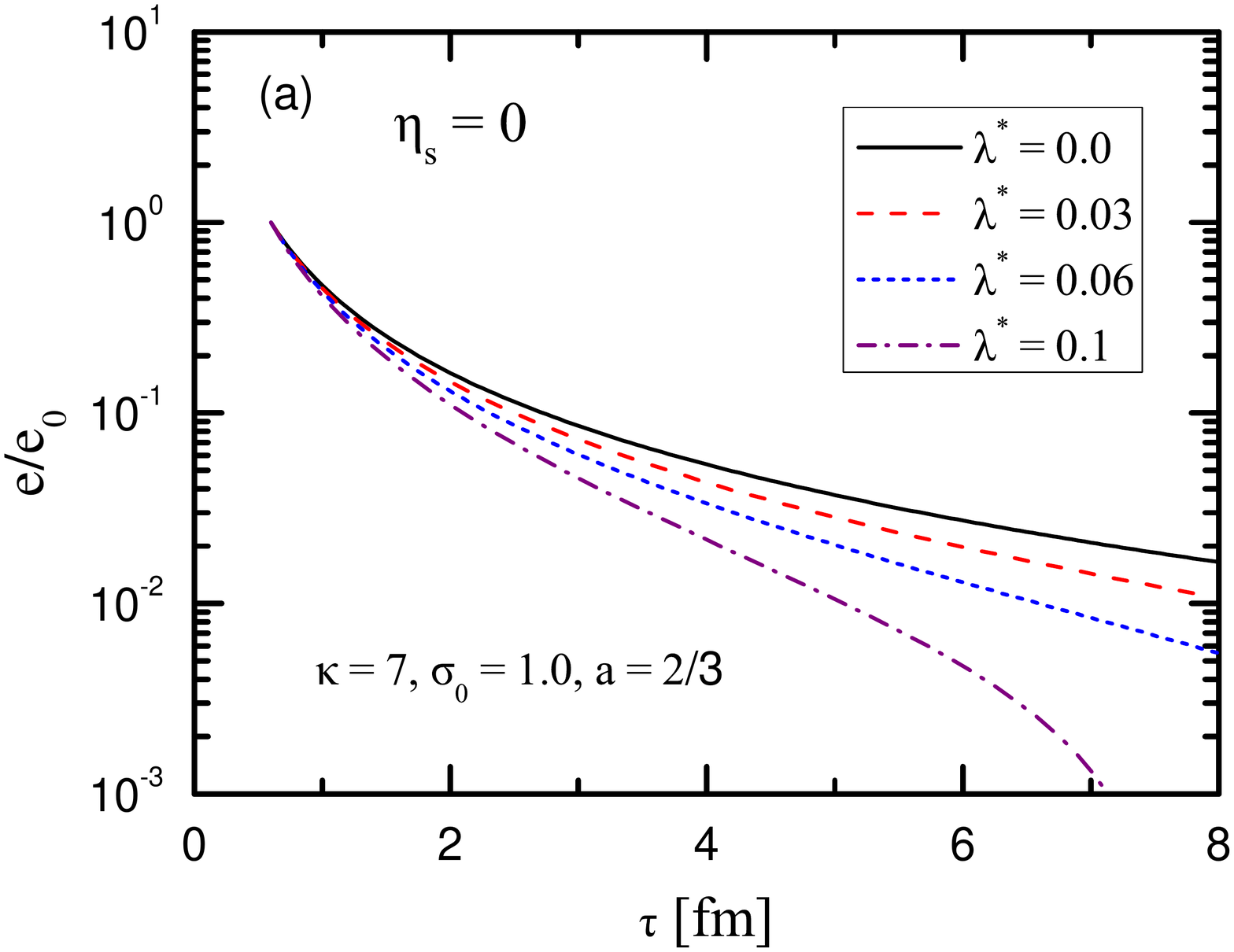,width=7.5cm}}
\end{minipage}
\hfill
\begin{minipage}[htb]{5cm}
\centerline{\epsfig{figure=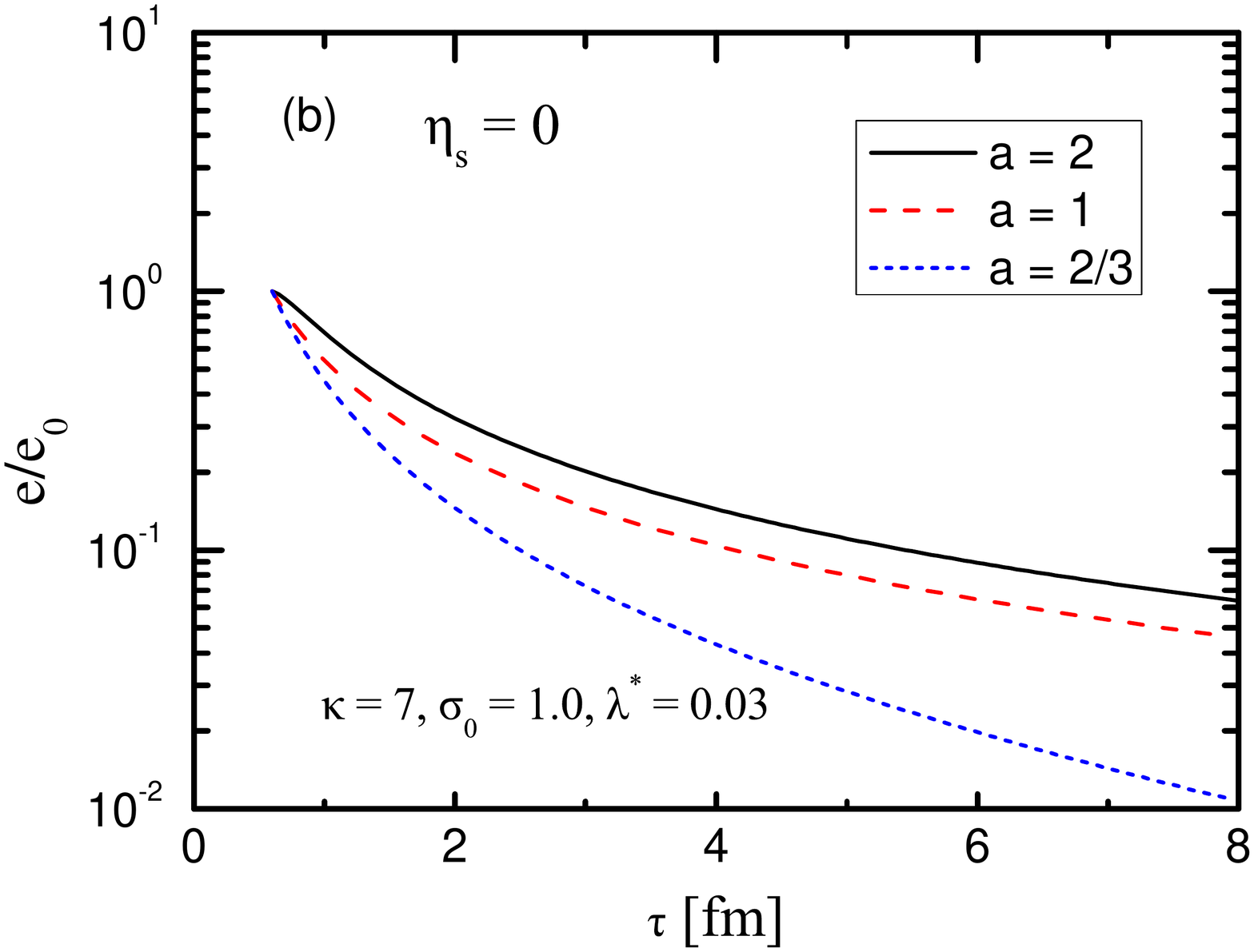,width=7.5cm}}
\end{minipage}
\hfill
\centering
\begin{minipage}[htb]{5cm}
\centerline{\epsfig{figure=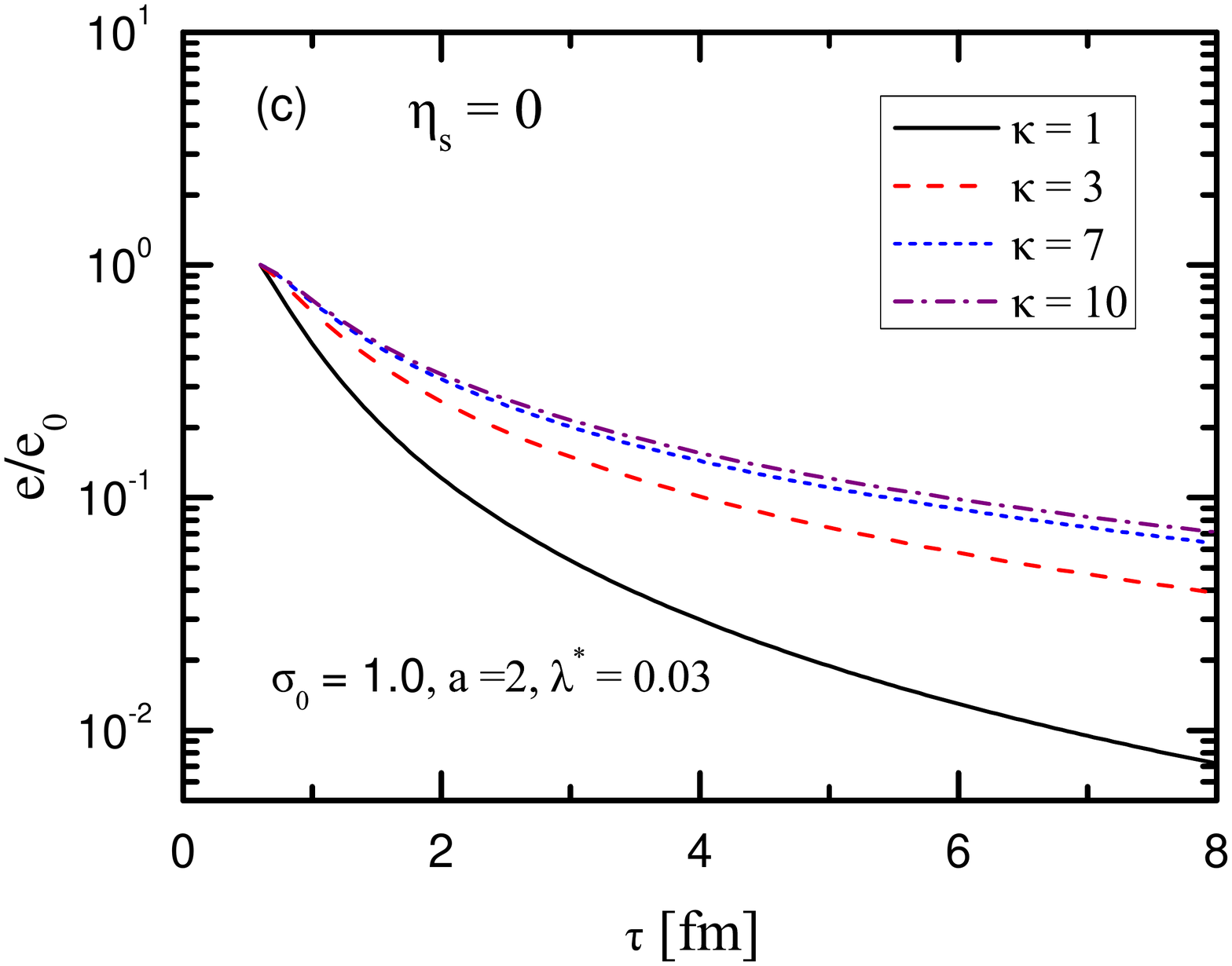,width=7.5cm}}
\end{minipage}
\caption{Evolution of fluid energy density $e/e_0$ as a function of proper time $\tau$ and we choose initial condition as $\widetilde{e}_{0}\equiv\widetilde{e}(\tau_{0})=1$. (a)Different lines refer to different levels of longitudinal acceleration parameter: $\lambda^{*}=0$ (black solid line), $\lambda^{*}=0.03$ (red dashed line), $\lambda^{*}=0.06$ (blue dotted line), $\lambda^{*}=0.1$ (magenta dot-dashed line). Clearly, it is gradually speed up the decay rate of fluid energy density with increasing acceleration parameter $\lambda^*$. (b)
Different lines refer to different levels of magnetic field decay parameter: $a=2$ (black solid line), $a=1$ (red dashed line), and $a=2/3$ (blue dotted line). Clearly, the fluid energy density decreases more rapidly for $a=2/3$ than in the case $a=1$, not to mention $a=2$. (c)Different lines refer to the evolution for $\kappa=1$ (black solid line), $\kappa=3$ (red dashed line), $\kappa=7$ (blue dotted line), and $\kappa=10$ (magenta dot-dashed line).
}
\label{fig2}
\end{figure}

\begin{figure}[htb]
\begin{minipage}[htb]{8.8cm}
\centerline{\epsfig{figure=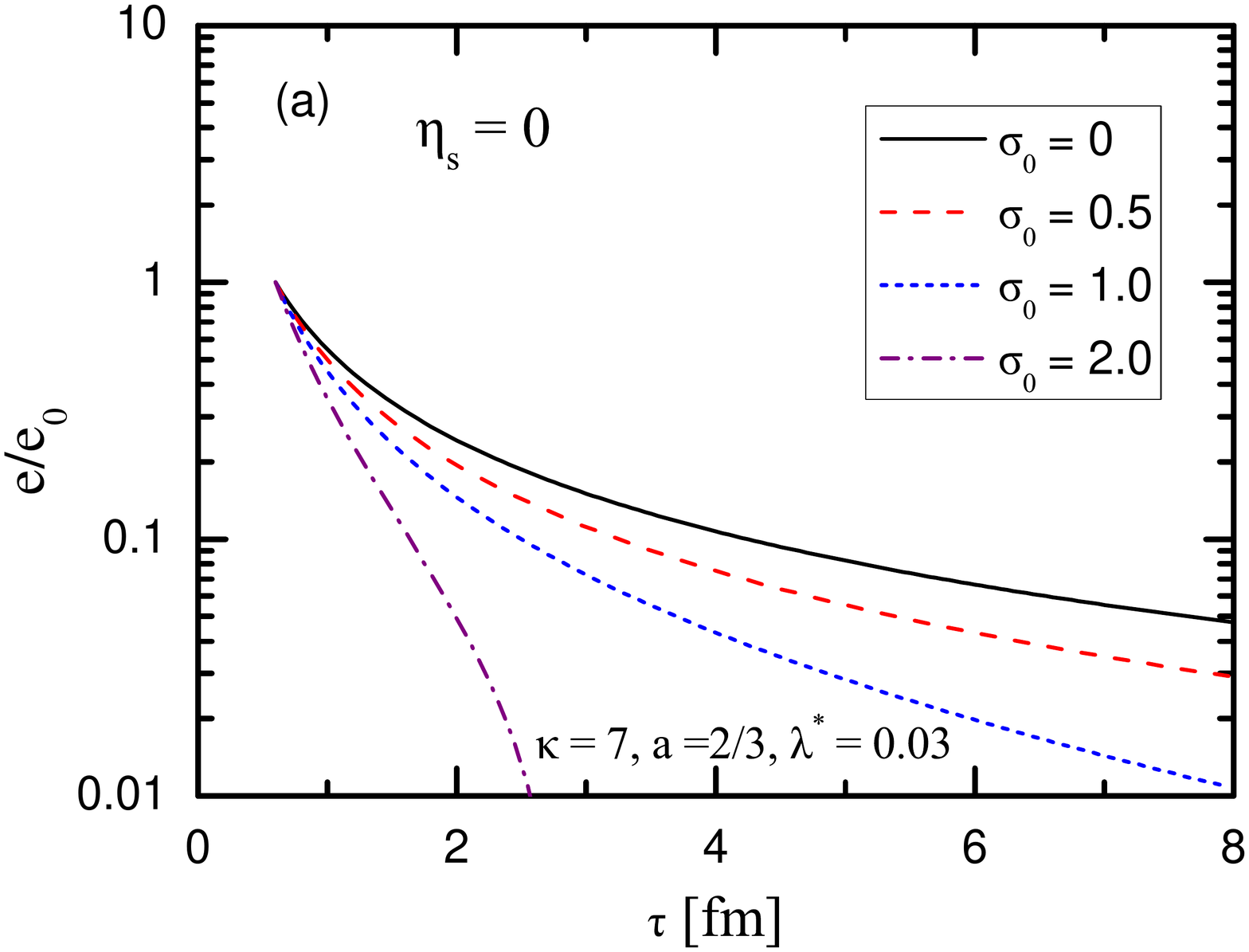,width=8.5cm}}
\end{minipage}
\hfill
\begin{minipage}[htb]{8.8cm}
\centerline{\epsfig{figure=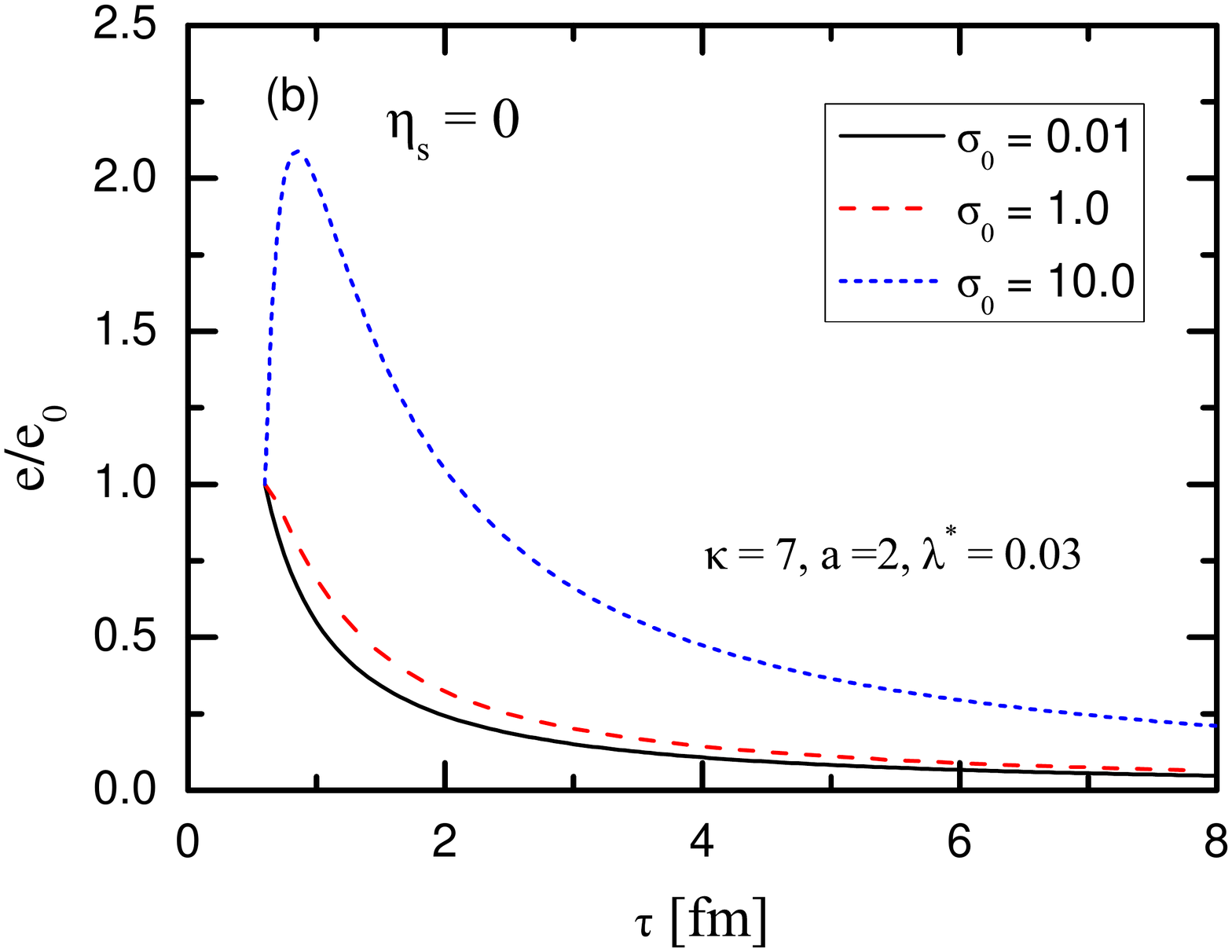,width=8.5cm}}
\end{minipage}
\hfill
\begin{minipage}[htb]{8.8cm}
\centerline{\epsfig{figure=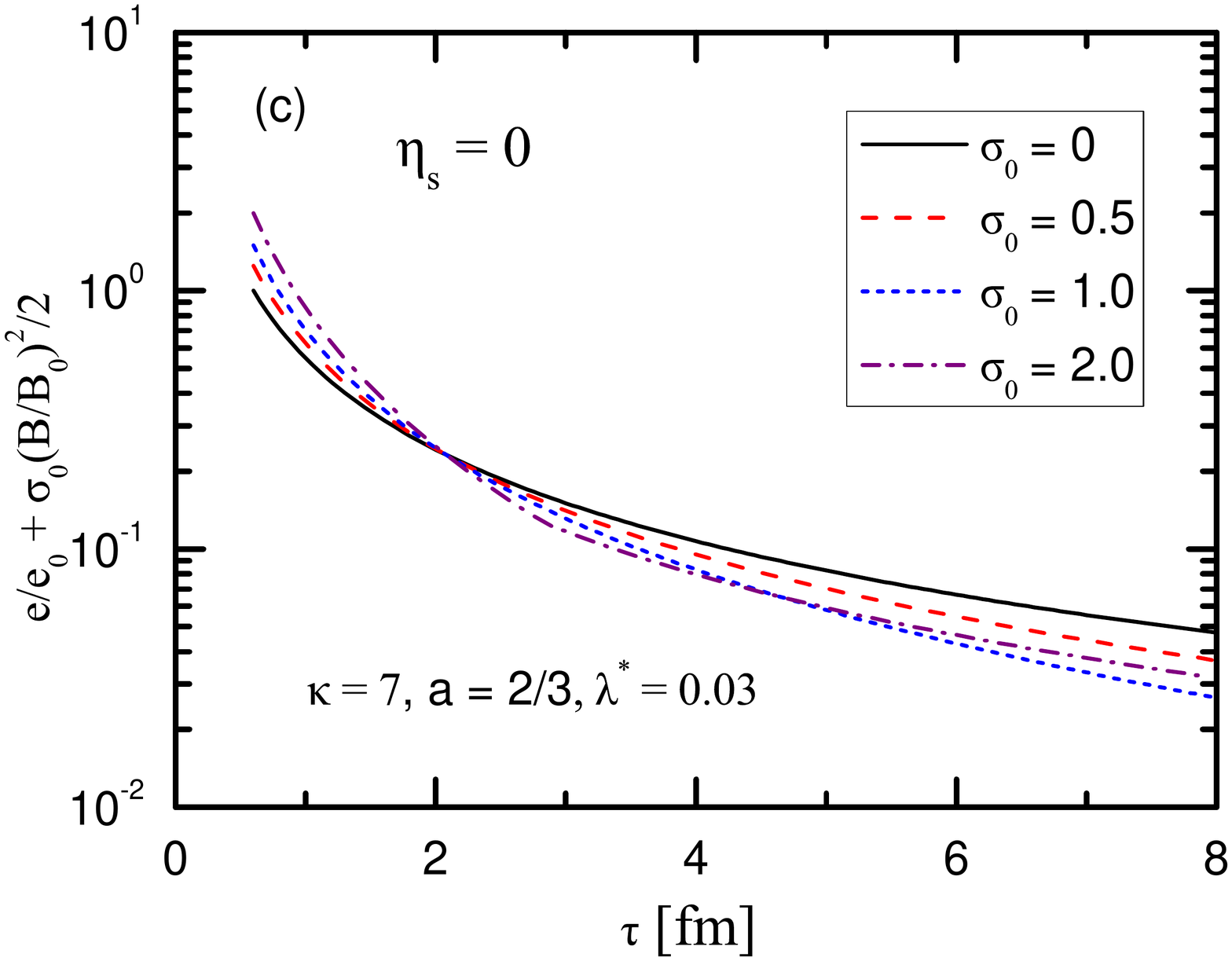,width=8.5cm}}
\end{minipage}
\hfill
\begin{minipage}[htb]{8.8cm}
\centerline{\epsfig{figure=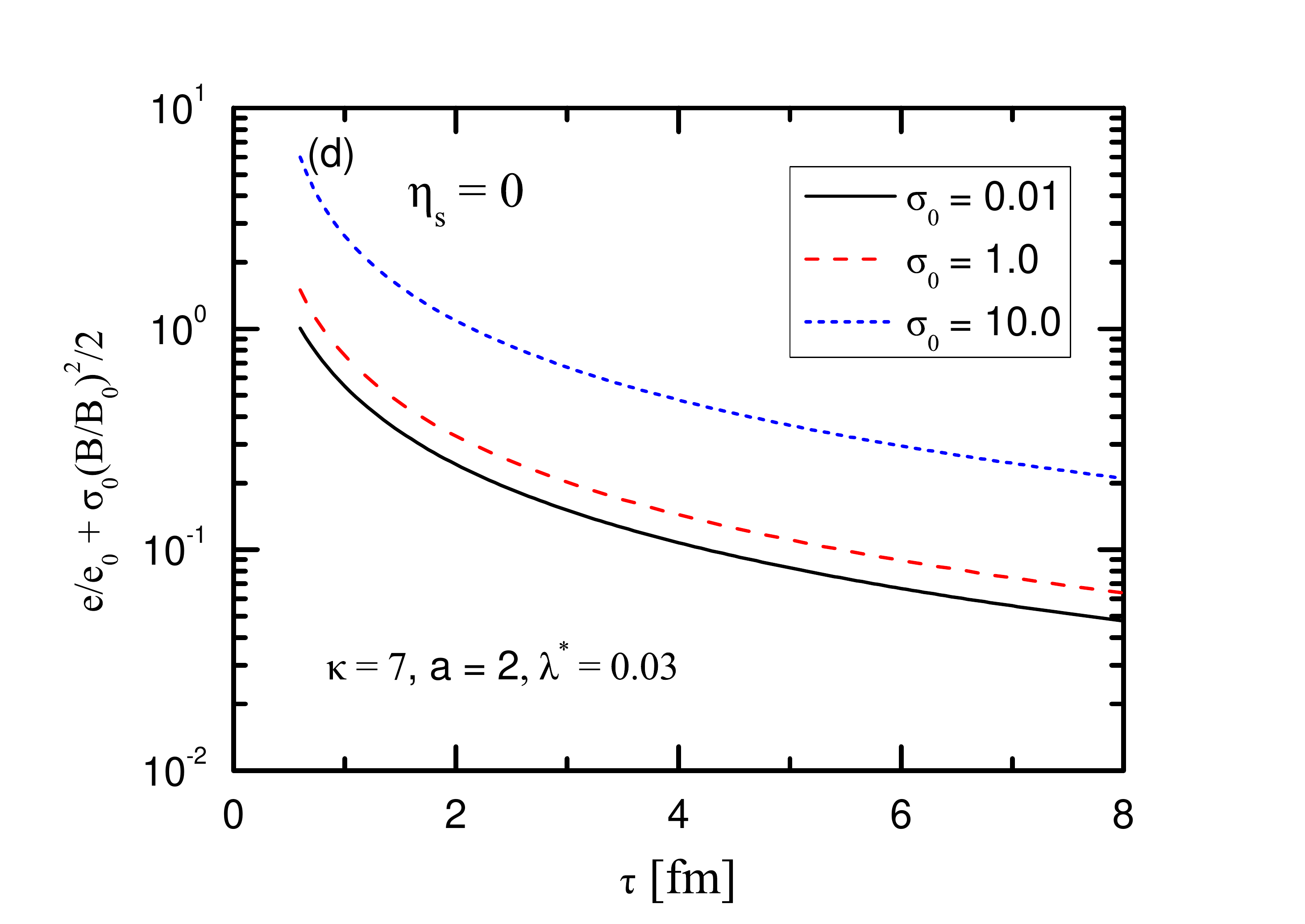,width=8.5cm}}
\end{minipage}
\caption{The evolution of the fluid energy density $e/e_0$ (upper panel) and the total energy density $e/e_{0}+\sigma_{0}(B/B_{0})^{2}/2$ (lower panel) in the different cases and when the parameters are set to $a=2/3$ (left panel) and $a=2$ (right panel). (Left panel) Different lines refer to different levels of the initial magnetization: $\sigma=0$ (black solid line), $\sigma=0.5$ (red dashed line), $\sigma=1.0$ (blue dotted line), and $\sigma=2.0$ (magenta dot-dashed line). (Right panel) Different lines refer to different levels of the initial magnetization, ranging from $\sigma_0=0.01$ (black solid line), $\sigma_0=1.0$ (red dashed line), and $\sigma_0=10.0$ (blue dotted line).}
\label{fig3}
\end{figure}

For the sake of comparison with Bjorken-Victor type flow,
we take the space-time rapidity $\eta_s=0$ in Fig.\ref{fig2} and Fig.\ref{fig3}. In Fig.\ref{fig2}, we compare the evolution of fluid energy density $\widetilde{e}$ for following different conditions: (\textbf{a})~different longitudinal acceleration parameter $\lambda^*$, (\textbf{b})~different magnetic field decay parameter $a$, and (\textbf{c})~different EoS parameter $\kappa$. In Fig.\ref{fig2} case (\textbf{a}), different lines represent to different values of the longitudinal acceleration parameter $\lambda^{*}$, ranging from $\lambda^{*}=0$ (Bjorken-Victor type flow without longitudinal acceleration effect; black solid line) up to cases with $\lambda^*=0.03$ (red dashed line), $\lambda^*=0.06$ (blue dotted line) and $\lambda^*=0.1$ (magenta dot-dashed line). In Fig.\ref{fig2} case (\textbf{b}), we show the evolution of the normalized energy density $\widetilde{e}$ for $a=2$ (black solid line), $a=1$ (ideal-MHD limit; red dashed line), and $a=2/3$ (blue dotted line). In Fig.\ref{fig2} case (\textbf{c}), different lines means different values of EoS $\kappa=1$ (CNC approximation; black solid line), $\kappa=3$ (red dashed line), $\kappa=7$ (blue dotted), and $\kappa=10$ (magenta dot-dashed line). As the graph illustrates, the longitudinal acceleration effect of the fluid increase the decay of the fluid energy density; $\widetilde{e}$ decays faster for $a=2/3$ than the ideal-MHD limit $a=1$ case, whereas for $a=2$ it initially decays more slowly and then decays asymptotically at the same rate as for the ideal-MHD $a=1$ case; the evolution of the fluid energy density $\widetilde{e}$ decays more quickly with decreasing $\kappa$.

In Fig.\ref{fig3}, we consider the evolution of the fluid energy density $\widetilde{e}$ (upper panel) and the total energy density $e/e_{0}+\sigma_{0}(B/B_{0})^{2}/2$ (lower panel) in the different cases and when the parameters are set to $a=2/3$ (left panel) and $a=2$ (right panel). Left panel report the evolution of fluid energy density and the total energy density $e/e_{0}+\sigma_{0}(B/B_{0})^{2}/2$ for $a=2/3$ and where different lines refer to different levels of the initial magnetization: $\sigma_0=0$ (black solid line), $\sigma_0=0.5$ (red dashed line), $\sigma_0=1.0$ (blue dotted line), and $\sigma_0=2$ (magneto dot-dashed line). It is clear that larger values of initial magnetization $\sigma_0$ will lead to a faster decrease in $\widetilde{e}$ and $e/e_{0}+\sigma_{0}(B/B_{0})^{2}/2$. Right panel shows the evolution of fluid energy density $\widetilde{e}$ and the total energy density $e/e_{0}+\sigma_{0}(B/B_{0})^{2}/2$ in the case $a=2$. In this case, different lines refer to different levels of the initial magnetization, $\sigma_0=0.01$ (black solid), $\sigma_0=1.0$ (red dashed), and $\sigma_0=10$ (blue dotted). As shown in the Fig.~\ref{fig3} (upper-right panel), it produces even a temporary increase in the fluid energy density evolution. This interesting phenomena, which can be associated with the resistive "heating up" of the fluid, and it depends on the values of the initial magnetization $\sigma_0$ and the magnetic field decay parameter $a$. This increase in the fluid energy density evolution will be larger for larger magnetic field decay parameter $a$ due to the fact that the Lorentz force allows energy to transfer back and forth between the magnetic field and the fluid. The total energy density of this system decays quickly with increasing $\sigma_0$ for $a<1$. Increasing $\sigma_0$ only adds energy density to the system, but does not alter the temporal evolution of the total energy density for the case with $a>1$.

\section{discussion and conclusions}
\label{section4}

We have investigated the evolution of the energy density of the QGP generated by the non-central heavy ion collisions by one-dimensional MHD flow in the limit of infinite electrical conductivity with longitudinal acceleration parameter $\lambda^*$ and got an exact solution under the CNC approximation. Compared with Bjorken-Victor type flow, the longitudinal acceleration effect accelerates the decay of the energy density. For larger $\kappa$ of EoS, the energy density decays more slowly, thus the temperature dependent EoS should be calculated from lattice QCD simulations.

Based on the definition of the acceleration coordinate (Rindler coordinate, Kottler-M\o ller coordinates, and Radar coordinates), the "acceleration parameter" $\lambda$ has following physics meaning: (1)$\lambda<0$, for heavy ion collisions, it means that the fireball system's element flowing into the fireball's core and the system's thermodynamics quantities density is increasing with the time, or in other words, the fireball system does not swell but contracts, which means after enough long time, there will creating a black holw; (2)$\lambda=0$ correspond to the rest fireball system; (3)$0<\lambda<1$, the fireball system's expansion speed is decelerating. The energy density deposit to large $\eta_s$; (4)$\lambda=1$, the fireball system's expansion speed is average; (5)$\lambda>1$, the fireball system's expansion is fast and many energy density deposit to the mid-rapidity $\eta_s$, which is consistent with the experimental data. Thus, we only focus on the case that longitudinal acceleration parameter $\lambda^*$ is greater than $0$ in the previous discussion.

For the case that the magnetic field evolution follows a power-law decay in proper time with exponent $a$, we find
the magnetic field decays more quickly than in the ideal-MHD case for $a>1$, while the magnetic field with $a<1$ correspond to a decay that is slower than in the ideal-MHD limit.
In heavy-ion collisions the remnants of colliding nuclei can give an additional contribution to the magnetic field to slow down its decay.
Thus, considering the case $a<1$ is reasonable in this paper.
It is clearly that larger values of initial magnetization $\sigma_0$ leads to faster decreasing in $\widetilde{e}$ for $a<1$. But it also results in a temporary increase in the fluid energy density for $a>1$. As we know, the magnetic field energy can be converted to fluid energy via Lorzent force, thus the evolution of the fluid energy density becomes more complex. For $a\rightarrow0$, the magnetic field is constant in proper time and does not evolve with the fluid.
Thus, the fluid energy density must decay very rapidly to keep this constant magnetic field.
For $a\rightarrow\infty$, the magnetic field decays fast and the energy is transferred to the fluid-element according to the energy-conservation law.
Thus, one can expect a peak of the energy density near the initial time, which is associated with a "reheating" of the fluid with longitudinal acceleration effect.

However, the recent estimates both from lattice QCD simulations ~\cite{Alessandro2013,Amato2013,Greif2014} and fitting of experimental data point~\cite{Yi Yin2014} toward high, but finite value for the electrical conductivity of the QGP. For a quantitative comparison with experimental data, the effects of the electrical resistivity has to be taken into account.

As a next step, we try to include the dissipative effects (shear and bulk viscosity and a finite electric conductivity), the rescatterings in the hadronic phase, the decays of hadronic resonance into stable hadrons and anomalous currents. Note it would be necessary to modify the Cooper-Frye formula by taking into account the presence of an electromagnetic field.

\begin{acknowledgements}
We specially thank Dirk H. Rischke for the useful suggestion about the MHD theory at the ATHIC2018. This work is in part supported by the Ministry of Science and Technology of China (MSTC) under the "973" Project No. 2015CB856904(4), by NSFC Grant Nos. 11735007, 11890711
This work was supported by the Sino-Hungarian bilateral cooperation program, under the Grand No.Te'T 12CN-1-2012-0016, by the financial supported from NNSF of China under grant No.11435004. Z-F. Jiang would like to thank T.~Cs\"org\H{o}, M.~Csan\'ad, L{\'e}vai P{\'e}ter and Gergely G{\'a}bor Barnafoldi for kind hospitality during his stay at Winger RCP, Budapest, Hungary.

\end{acknowledgements}

\end{document}